\documentclass[12pt]{article}

\def\d{\partial}

\begin{document}

\title{The nature of spacetime singularities}

\author{Alan D. Rendall\\
Max-Planck-Institut f\"ur Gravitationsphysik\\Albert-Einstein-Institut
\\ Am M\"uhlenberg 1\\
14476 Golm, Germany}

\date{}

\maketitle

\begin{abstract}
Present knowledge about the nature of spacetime singularities in the
context of classical general relativity is surveyed. The status of the 
BKL picture of cosmological singularities and its relevance to the cosmic
censorship hypothesis are discussed. It is shown how insights on cosmic 
censorship also arise in connection with the idea of weak null singularities 
inside black holes. Other topics covered include matter singularities and 
critical collapse. Remarks are made on possible future directions in research 
on spacetime singularities.
\end{abstract}

\section{Introduction}

The issue of spacetime singularities arose very early in the history
of general relativity and it seems that Einstein himself had an
ambiguous relationship to singularities. A useful source of information
on the confusion surrounding the subject in the first half century of
general relativity is \cite{earman}. The present article 
is a survey of the understanding we have of spacetime singularities
today.

Before concentrating on general relativity, it is useful to think
more generally about the concept of a singularity in a physical
theory. In the following the emphasis is on classical field theories
although some of the discussion may be of relevance to quantum theory
as well. When a physical system is modelled within a classical
field theory, solutions of the field equations are considered.
If it happens that at some time physically relevant quantities
become infinite at some point of space then we say that there is
a singularity. Since the physical theory ceases to make sense when
basic quantities become infinite a singularity is a sign that 
the theory has been applied beyond its domain of validity. To get a 
better description a theory of wider applicability should be used. 
Note that the occurrence of singularities does not say that a theory
is bad - it only sets limits on the domain of physical phenomena 
where it can be applied.

In fact almost any field theory allows solutions with singularities
if attention is not restricted to those solutions which are likely to 
be physically relevant. In this context a useful criterion is 
provided by the specification of solutions by initial data.
This means that we only consider solutions which have the property
that there is some time at which they contain no singularities.
Then any singularities which occur must be the result of a dynamical
evolution. With this motivation, singularities will be discussed in
the following in the context of the initial value problem. Only those
singularities are considered which develop from regular initial 
configurations. This has the consequence that linear field theories,
such as source-free Maxwell theory, are free of singularities.

In the case of the Einstein equations, the basic equations of general
relativity, the notion of singularity becomes more complicated due
to the following fact. A solution of the Einstein equations consists 
not just of the spacetime metric, which describes the gravitational
field and the geometry of spacetime, but also the spacetime manifold 
on which the metric is defined. In the case of a field theory in
Newtonian physics or special relativity we can say that a solution
becomes singular at certain points of spacetime, where the basic physical 
quantities are not defined. Each of these points can be called a
singularity. On the other hand, a singularity in general relativity
cannot be a point of spacetime, since by definition the spacetime 
structure would not be defined there.

In general relativity the wordline of a free particle is described by
a curve in spacetime which is a timelike or null geodesic, for a massive
or massless particle respectively. There is also a natural class of time 
parameters along such a geodesic which, in the timelike case, coincide
up to a choice of origin and a rescaling with the proper time in the 
rest frame of the particle. If the worldline of a particle only exists
for a finite time then clearly something has gone seriously wrong.
Mathematically this is called geodesic incompleteness. A spacetime
which is a solution of the Einstein equations is said to be {\it singular} if
it is timelike or null geodesically incomplete. Informally we say in this 
case that the spacetime \lq contains a singularity\rq\ but the definition 
does not include a description of what a \lq singularity\rq\ or \lq singular 
point\rq\ is. There have been attempts to define ideal points 
which could be added to spacetime to define a mathematical boundary
representing singularities but these have had limited success.

When working practically with solutions of the Einstein equations it
is necessary to choose coordinates or other similar auxiliary objects
in order to have a concrete description. In general relativity we are free 
to use any coordinate system and this leads to a problem when considering
singularities. Suppose that a metric written in coordinates is such
that the components of the metric become infinite as certain values of
the coordinates are approached. This could be a sign that there is
a spacetime singularity but it could also simply mean that those
coordinates break down at some points of a perfectly regular solution.
This might be confirmed by transforming to new coordinates where the 
metric components have a regular extension through the apparent 
singularities. A way of detecting singularities within a coordinate
system is to find that curvature invariants become infinite. These
are scalar quantities which measure the curvature of spacetime and
if they become infinite this is a sure sign that a region of spacetime
cannot be extended. It is still not completely clear what is happening 
since the singular values of the coordinates might correspond to
singular behaviour in the sense of geodesic incompleteness or they 
might be infinitely far away.

A breakthrough in the understanding of spacetime singularities was
the singularity theorem of Penrose \cite{penrose1} which identified general
conditions under which a spacetime must be geodesically incomplete.
This was then generalized to other situations by Hawking and others.
The singularity theorems are proved by contradiction. Their strength
is that the hypotheses required are very general and their weakness
is that they give very little information about what actually happens
dynamically. If the wordline of a particle ceases to exist after finite
proper time then it is reasonable to ask for an explanation, why the
particle ceased to exist. It is to be expected that some extreme physical
conditions play a role. For instance, the matter density or the curvature, 
representing tidal forces acting on the particle,  becomes unboundely 
large. From this point of view one would like to know that curvature
invariants become unbounded along the incomplete timelike or null 
geodesics. The singularity theorems give no information on this
question which is that of the nature of spacetime singularities.
The purpose of the following is to explain what is known about this
difficult question.

The hypotheses of the singularity theorems do not include very stringent
assumptions about the matter content of spacetime. All that is needed
are certain inequalities on the energy-momentum tensor $T_{\alpha\beta}$, 
the energy conditions \cite{he}. Let $V^\alpha$ and $W^\alpha$ be arbitrary 
future pointing timelike vectors. The {\it dominant energy condition} is that 
$T_{\alpha\beta}V^\alpha W^\beta\ge 0$. The {\it strong energy condition} is, 
provided the cosmological constant is zero, equivalent to the condition that 
$R_{\alpha\beta}V^\alpha V^\beta\ge 0$ where $R_{\alpha\beta}$ is the Ricci
tensor. The {\it weak energy condition} is
that $T_{\alpha\beta}V^\alpha V^\beta\ge 0$. The weak energy condition has
the simple physical interpretation that the energy density of matter is
non-negative in any frame of reference. The vector $V^\alpha$ is the
four-velocity of an observer at rest in that frame of reference. It is not 
reasonable to expect that the nature of spacetime singularities can be 
determined on the basis of energy conditions alone - more detailed 
assumptions on the matter content are necessary.

It follows from the above discussion that spacetime singularities
should be associated with reaching the limits of the physical
validity of general relativity. Quantum effects can be expected to
come in. If this is so then to go further the theory should be replaced by 
some kind of theory of quantum gravity. Up to now we have no definitive theory
of this kind and so it is not clear how to proceed. The strategy
to be discussed in the following is to work entirely within classical
general relativity and see what can be discovered. It is to be
hoped that this will provide useful input for the future investigation
of singularities within a more general context. The existing attempts to
study the question of singularities within different approaches to
quantum gravity, including the popular idea that quantum gravity should
eliminate the singularities of classical general relativity, will not be 
discussed here. For a discussion of one direction where progress is 
being made, see the article of M. Bojowald in this volume. 

A key question about singularities in general relativity is whether they
are a disaster for the theory. If a singularity can be formed and then 
influence the evolution of spacetime then this means a breakdown of
predictability for the theory. For we cannot (at least within the classical
theory) predict anything about the influence a singularity will have. A
singularity which can causally influence parts of spacetime is called a
naked singularity. It is important for the predictive power of general
relativity that naked singularities be ruled out. This has been 
formulated more precisely by Penrose as the cosmic censorship hypothesis
\cite{penrose2}, \cite{penrose3}. In fact there are two variants of this,
weak and strong cosmic censorhip. Despite the names neither of these implies
the other \cite{wald}. Proving the cosmic censorship hypothesis is one of 
the central mathematical problems of general relativity. In fact the task 
of finding the right formulation of the conjecture is already a delicate one. 
It is necessary to make a genericity assumption and to restrict the matter 
fields allowed. More details on this are given in later sections.

One of the most important kinds of singularity in general relativity
is the initial cosmological singularity, the big bang. The structure of
cosmological singularities is the subject of section \ref{cosm}. Another
important kind of singularity is that inside black holes. The recent
evolution of ideas about the internal structure of black holes is discussed
in section \ref{bh}. An important complication in the study of singularities
resulting from the properties of gravity is that they may be obscured by
singularities due to the description of matter. This is the theme of section
\ref{matter}. In section \ref{critical} singularities are discussed which
arise at the threshhold of black hole formation and which are still quite
mysterious. Section \ref{conclusion} takes a cautious look at the future
of research on spacetime singularities.

\section{Cosmological singularities}\label{cosm}

The simplest cosmological models are those which are homogeneous and 
isotropic, the FLRW (Friedmann-Lema\^itre-Robertson-Walker) models
with some suitable choice of matter model such as a perfect fluid.
In this context it is seen that the energy density blows up at
some time in the past. An early question was whether this singularity
might be an artefact of the high symmetry. The intuitive idea is that if 
matter collapses in such a way that particles are aimed so as to all
end up at the same place at the same time there will be a singularity.
On the other hand if this situation is perturbed so that the particles 
miss each other the singularity might be removed. On the basis of 
heuristic arguments, Lifshitz and Khalatnikov \cite{lk} suggested that
for a generic perturbation of a FLRW model there would be no singularity.
We now know this to be incorrect. This work nevertheless led to a very
valuable development of ideas in the work of Belinskii, Khalatnikov and 
Lifshitz \cite{bkl1}, \cite{bkl2} 
which is one of the main sources for our present picture of
cosmological singularities.

What was the problem with the original analysis? An ansatz was made for the
form of the metric near the singularity and it was investigated how
many free functions can be accomodated in a certain formal expansion.
It was found that there was one function less than there is in the general
solution of the Einstein equations. It was concluded that the most
general solution could not have a singularity. This shows us something 
about the strengths and weaknesses of heuristic arguments. These are
limited by the range of possibilities that have occurred to those producing
the heuristics. Nevertheless they may, in expert hands, be the most efficient
way of getting nearer to the truth. 

It was the singularity theorems, particularly the Hawking singularity 
theorem, which provided convincing evidence that cosmological singularities
do occur for very large classes of initial data. In particular they showed 
that the presence of a singularity (in the sense of geodesic incompleteness)
is a stable property under small perturbations of the FLRW model. Thus a 
rigorous mathematical theorem led to progress in our understanding of 
physics. The use of mathematical theorems is very appropriate because the
phenomena being discussed are very far from most of our experience of 
the physical world and so relying on physical intuition alone is dangerous.  

The singularity theorems give almost no information on the nature of the 
singularities. In order to go further it makes sense to attempt to 
combine rigorous mathematics, heuristic arguments and numerical
calculations and this has led to considerable progress.

The picture developed by Belinskii, Khalatnikov and Lifshitz (BKL) has
several important elements. These are:

\begin{itemize}
\item Near the singularity the evolution of the geometry at different 
spatial points decouples so that the solutions of the partial differential 
equations can be approximated by solutions of ordinary differential 
equations.
\item For most types of matter the effect of the matter fields on the 
dynamics of the geometry becomes negligible near the singularity
\item The ordinary differential equations which describe the asymptotics 
are those which come from a class of spatially homogeneous solutions
which constitute the mixmaster model. General solutions of these equations
show complicated oscillatory behaviour near the singularity.
\end{itemize}

\noindent
The first point is very surprising but a variety of analytical and 
numerical studies appear to support its validity. The extent to which the 
above points have been confirmed will now be discussed.

The mixmaster model is described by ordinary differential equations and
so it is a huge simplification compared to the full problem. Nevertheless
even ordinary differential equations can be very difficult to analyse.
The solutions show complicated behaviour in the approach to the singularity
and this is often called chaotic. This description is somewhat problematic 
since many of the usual concepts for defining chaos are not applicable. This 
point will not be discussed further here. For many years the oscillations in
solutions of the mixmaster model were studied by heuristic and numerical
techniques. This led to a consistent picture but turning this picture 
into mathematical theorems was an elusive goal. Finally this was achieved
in the work of Ringstr\"om \cite{ringstrom1} so that the fundamental 
properties of the mixmaster model are now mathematically established.

With the mixmaster model under control, the next obvious step in 
confirming the BKL picture would be to show that it serves as
a template for the behaviour of general solutions near the singularity.
The work of BKL did this on a heuristic level. Attempts to recover
their conclusions in numerical calculations culminated in the work
of Garfinkle \cite{garfinkle}. Previously numerical investigations of the
question had been done under various symmetry assumptions. Solutions 
without symmetry were handled for the first time in \cite{garfinkle}.
On the analytical side things do not look so good. There is not a
single case with both inhomogeneity and mixmaster oscillations which
has been analysed rigorously and this represents an outstanding challenge.
One possible reason why it is so difficult will be described below.

One of the parts of the BKL picture contains the qualification \lq for
most types of matter\rq. There are exceptional types of matter where
things are different. A simple example is a massless linear scalar
field. It was already shown in \cite{belinskii1} that in the presence 
of a scalar field the BKL analysis leads to different conclusions. It
is still true that the dynamics at different spatial points decouples
but the evolution is such that important physical quantities are 
ultimately monotone instead of being oscillatory as the singularity
is approached. In this case it has been possible to obtain a mathematical
confirmation of the BKL picture. In \cite{andersson} it was shown that
there are solutions of the Einstein equations coupled to a scalar field
which depend on the maximal number of free functions and which have
the asymptotic behaviour near the singularity predicted by the BKL
picture.

As a side remark, note that in many string theory models there is
a scalar field, the dilaton, which might kill mixmaster oscillations.
Also, a BKL analysis of the vacuum Einstein equations in higher
dimensions shows that the oscillations of generic solutions vanish
when the spacetime dimension is at least eleven \cite{henneaux} and 
string theory leads to the consideration of models of dimension greater
than four. So could mixmaster oscillations be eliminated in low energy 
string theory? An investigation in \cite{damour} shows that they are not.
The simplifying effect of the dilaton and the high dimension is
prevented by other form fields occurring in string theory. With
certain values of the coupling constants in  field theories of the type 
coming up in low energy string theory there is monotone behaviour near 
the singularity and theorems can be proved \cite{dhrw}. However the
work of \cite{damour} shows that these do not include the values of
the coupling constants coming from the string theories which are now 
standard.

A feature which makes oscillations so difficult to handle is that they
are in general accompanied by large spatial gradients. Consider some
physical quantity $f(t,x)$ in the BKL picture in a case without
oscillations. Then it is typical that quantities like $\d_i f/f$, where
the derivatives are spatial derivatives, remain bounded near the singularity.
However it can happen that this is only true for most spatial points and
that there are exceptional spatial points where it fails. In a situation
of mixmaster type where there are infinitely many oscillations as the 
singularity is approached the BKL picture predicts that there will be 
more and more exceptional points without limit as the singularity is
approached. It has even been suggested by Belinskii that this shows
that the original BKL assumptions are not self-consistent \cite{belinskii2}.
In any case, it seems that the question, in what sense the BKL picture
provides a description of cosmological singularities, is a subtle one.

Large spatial gradients can also occur in solutions where the evolution
is monotone near the singularity. It can happen that before the 
monotone stage is reached there are finitely many oscillations and
that these produce a finite number of exceptional points. In the context
of Gowdy spacetimes this has been shown rather explicitly. The features
with large spatial gradients (spikes) were discovered in numerical work
\cite{berger} and later captured analytically \cite{weaver}. This allowed
the behaviour of the curvature near the singularity to be determined.

An important issue to be investigated concerning cosmological singularities
is that of cosmic censorship. In this context it is strong cosmic censorship
which is of relevance and a convenient mathematical formulation in terms of
the initial value problem has been given by Eardley and 
Moncrief \cite{eardley}. To any initial data set for the Einstein equations
there exists a corresponding maximal Cauchy development. (For background on
the initial value problem for the Einstein equations see \cite{friedrich}.)
The condition that a spacetime is uniquely determined by initial data is
global hyperbolicity. The maximal Cauchy development is in a well-defined 
sense the largest globally hyperbolic spacetime with the chosen initial 
data. It may happen that the maximal Cauchy development can be extended to
a larger spacetime, which is then of course no longer globally 
hyperbolic. The boundary of the initial spacetime in the extension is
called the Cauchy horizon. The extended spacetime can no longer be uniquely 
specified by initial data and this corresponds physically to a breakdown of 
predictability. A famous example where this happens is the Taub-NUT spacetime 
\cite{he}. This is a highly symmetric solution of the Einstein vacuum 
equations. The extension which is no longer globally hyperbolic contains 
closed timelike curves.

How can the existence of the Taub-NUT and similar spacetimes be reconciled 
with strong cosmic censorship? A way to do this would be to show that
this behaviour only occurs for exceptional initial data and that for
generic data the maximal globally hyperbolic development is inextendible.
This has up to now only been achieved in the simplified context of classes
of spacetimes with symmetry. These classes of spacetimes are not generic
and so they do not directly say anything about cosmic censorship. However
they provide model problems where more can be learned about the conceptual
and technical issues which arise in trying to prove cosmic censorship.
This kind of restricted cosmic censorship has been shown for many spatially
homogeneous spacetimes in \cite{chrusciel} and \cite{rendallhom} and for plane
symmetric solutions of the Einstein equations coupled to a massless scalar
field \cite{tegankong}. The most general, and most remarkable, result of this
kind up to now is the proof by Ringstr\"om \cite{ringstrom3} of strong cosmic
censorship restricted to the class of Gowdy spacetimes. He shows that all
the solutions in this class of inhomogeneous vacuum spacetimes with
symmetry are geodesically complete in the future \cite{ringstrom2} and that
for generic initial data the Kretschmann scalar 
$R^{\alpha\beta\gamma\delta}R_{\alpha\beta\gamma\delta}$ tends to infinity
uniformly as the singularity is approached. Major difficulties in doing
this are the fact that there do exist spacetimes in this class where the 
maximal Cauchy development is extendible and that spikes lead to great 
technical complications. Roughly speaking, Ringstr\"om shows under a 
genericity assumption that the most complicated thing that can happen in the 
approach to the singularity is that there are finitely many spikes of the 
kind constructed in \cite{weaver}.   

Another kind of partial result is to show that an expanding cosmological
spacetime is future geodesically complete. This can be interpreted as
saying that any singularities must lie in the past. There is up to now
just one example of this for spacetimes not required to satisfy any symmetry
assumptions. This is the work of Andersson and Moncrief \cite{am} where
they show that any small but finite vacuum perturbation of the initial
data for the Milne model has a maximal Cauchy development which is future
geodesically complete.    

Already in the class of homogeneous and isotropic spacetimes there are
models with an initial singularity which recollapse and have a second
singularity in the future. Not much is known about general criteria for
recollapse. The closed universe recollapse conjecture \cite{bg} says
that any spacetime with a certain type of topology (admitting a metric
of positive scalar curvature) and satisfying the dominant and strong 
energy conditions must recollapse. No counterexample is known but the
conjecture has only been proved in cases with high symmetry 
\cite{burnett1}, \cite{burnett2}. 

\section{Black hole singularities}\label{bh}

One of the most famous singular solutions of the Einstein equations is 
the Schwarzschild solution representing a spherical black hole. There
is a singularity inside the black hole where the Kretschmann scalar
diverges uniformly. It looks very much like a cosmological 
singularity. The singularity is not visible to far away observers.
The points of spacetime from which no future-directed causal geodesic
can escape to infinity constitute by definition the black hole region and 
its boundary is the event horizon. The situation in the Schwarzschild 
solution can be described informally by saying that the singularity is 
covered by an event horizon. The idea of weak cosmic censorship, a
concept which will not be precisely defined here, is that any singularity
which arises in gravitational collapse is covered by an event horizon.
For more details see \cite{wald}, \cite{christodoulou1}.

The central question which is to be answered is what properties of the
Schwarzschild solution are preserved under perturbations of the initial
data. Christodoulou has studied the spherical gravitational collapse of 
a scalar field in great detail \cite{christodoulou2}. Among his results
are the following. There are initial data leading to the formation of
naked singularities but for generic initial data this does not happen.
The structure of the singularity has been analysed and it shows strong
similarities to the Schwarzschild case.

A key concept in the Penrose singularity theorem is that of a {\it trapped
surface}. It has been shown by Dafermos \cite{dafermos2} that some of the 
results of Christodoulou can be extended to much more general spherically
symmetric spacetimes under the assumptions that there exists at least
one trapped surface and that the matter fields present are well-behaved
in a certain sense. They should not form singularities outside the 
black hole region. This condition on the matter fields was verified
for collisionless matter in \cite{dafermosr}. The fact that it is
satisfied for certain non-linear scalar fields led to valuable insights
in the discussion of the formation of naked singularities in a class
of models motivated by string theory \cite{horowitz}, \cite{dafermoss}.

When the Schwarzschild solution is generalized to include charge or
rotation the picture changes dramatically. In the relevant solutions,
the Reissner-Nordstr\"om and Kerr solutions, the Schwarzschild singularity
is replaced by a Cauchy horizon. At one time it was hoped that this was
an artefact of high symmetry and that a further perturbation would turn
it back into a curvature singularity. There was also a suggested mechanism
for this, namely that radiation coming from the outside would undergo an
unlimited blue shift as it approached the potential Cauchy horizon. Things
turned out to be more complicated, as discovered by Poisson and Israel
\cite{poisson}.

The new picture in \cite{poisson} for a perturbed charged black hole was
that the Cauchy horizon, where the metric is smooth, would be replaced
by a null hypersurface where, although the metric remains continuous and 
non-degenerate, the curvature blows up. They called this a weak null
singularity. The heuristic work of \cite{poisson} was followed up by
numerical work \cite{piran} and was finally turned into rigorous mathematics
by Dafermos \cite{dafermos1}. Perhaps the greatest significance of this
work on charged black holes is its role as a model for rotating black
holes. For the more difficult case of rotation much less is known although
there is some heuristic analysis \cite{ori}. At this point it is appropriate
to make a comment on heuristic work which follows on from remarks in the
last section. For several years it was believed, on the basis of a heuristic
analysis in \cite{moss}, that a positive cosmological constant would 
stabilize the Reissner-Nordstr\"om Cauchy horizon. This turned out, however,
to be another case where not all relevant mechanisms had been thought of.
In a later heuristic analysis \cite{brady} it was pointed out that there is 
another instability mechanism at work which reverses the conclusion.  

The case of weak null singularities draws attention to an ambiguity in
the definition of strong cosmic censorship. The formulation uses the
concept of extension of a spacetime. To have a precise statement is must
be specified how smooth a geometry must be in order to count as an 
extension. This may seem at first sight like hair splitting but in the case
of weak null singularities the answer to the question of strong cosmic
censorship is quite different depending on whether the extension is 
required to be merely continuous or continuously differentiable. A
related question is whether the extension should be required to satisfy
the Einstein equations in some sense. 

A question which does not seem to have been investigated is that of
the consistency of weak null singularities with the BKL picture. It
is typical to study black holes in the context of isolated systems. 
In reality we expect that black holes form in cosmological models
which expand for ever. Do such 'cosmological black holes' show the
same features in their interior as asymptotically flat ones? If so
then this would indicate the existence of large classes of cosmological
models whose singularities do not fit into the BKL picture. (It was 
never claimed that this picture must apply to all cosmological 
singularities.) A major difficulty in investigating this issue is that
the class of solutions of the Einstein equations of interest does not
seem to be consistent with any symmetry assumptions. A related question
is that of the relationship between weak cosmic censorship, which is
formulated in asymptotically flat spacetimes, and strong cosmic censorship,
which makes sense in a cosmological context.

There are important results showing that no black holes form under
certain circumstances. In the fundamental work of Christodoulou and 
Klainerman \cite{ck} it was shown that small asymptotically flat
data for the Einstein vacuum equations lead to geodesically complete
spacetimes. See also \cite{lr}. 

\section{Shells and shocks}\label{matter}

A serious obstacle to determining the structure of spacetime singularites
is that many common matter models develop singularities in flat space.
This is in particular the case for matter models which are phenomenological
rather than coming directly from fundamental physics. These matter models,
when coupled to the Einstein equations, must be expected to lead to 
singularities which have little to do with gravitation which we may call
matter singularities. These singularities are just a nuisance when we want to
study spacetime singularities as fundamental properties of Einstein gravity.

There has been much study of the Einstein equations coupled to dust. It
is not clear that they teach us much. In flat space dust forms shell-crossing
singularities where a finite mass of dust particles end up at the same 
place at the same time. The density blows up there. In curved space this
leads to naked singularities \cite{yodzis}. These occur away from the centre
in spherical symmetry. Finite time breakdown of self-gravitating dust can 
also be observed in cosmological spacetimes \cite{rendalldust}. This shows the
need for restricting the class of matter considered if a correct formulation
of cosmic censorship is to be found. In a more realistic perfect fluid the 
pressure would be expected to eliminate these singularities. On the other hand 
it is to be expected that shocks form, as is well-known in flat space. The 
breakdown of smoothness in self-gravitating fluids with
pressure was proved in \cite{stahl}. At this point we must once again 
confront the question of what is a valid extension. In some cases solutions
with fluid may be extended beyond the time when the classical solution 
breaks down \cite{lefloch}. The extended solution is such that the basic
fluid variables are bounded but their first derivatives are not. The 
uniqueness of these solutions in terms of their initial data is not known
but uniqueness results have recently been obtained in the flat space
case \cite{bressan}.

A matter model which is better behaved than a fluid is collisionless matter
described by the Vlasov equation. It forms no singularities in flat space
and there are various cases known where self-gravitating collisionless
matter can be proved to form no singularities. For instance this is the case 
for small spherically symmetric asymptotically flat initial data \cite{rr}. 
There is no case known where collisionless matter does form a matter 
singularity. Also in spherical symmetry it never forms a singularity away 
from the centre so that the analogue of shell-crossing singularities is ruled 
out \cite{rrs}. In view of the investigations up to now collisionless matter 
seems to be as well-behaved as vacuum with respect to the formation of 
singularities.

\section{Critical collapse}\label{critical}

Evidence for a new kind of singularity in gravitational collapse was
discovered by Choptuik \cite{choptuik}. His original work concerned the
spherically symmetric collapse of a massless scalar field but it has
been extended in many directions since then. The basic idea is as follows.
For small initial data the corresponding solution disperses leaving 
behind flat space. For very large data a black hole is formed. If
a one-parameter family of data is taken interpolating between these two
extreme cases what happens to the evolutions for intermediate values of
the parameter? It is found that there is a unique parameter value (the 
critical value) separating the two regimes and that near the critical 
value the solutions show interesting, more or less universal, behaviour.
The study of these phenomena is now known under the name of critical
collapse.

Most of the work which has been done on critical collapse is numerical.
There is a heuristic picture involving dynamical systems which is useful
in predicting certain features of the results of numerical calculations.
Up to now there are no rigorous results on critical phenomena. It is 
interesting to note that at least some of the features of critical
collapse are not unique to gravity and may be seen in many systems of
partial differential equations \cite{bizon}.

The results on critical collapse indicate the occurrence of a class
of naked singularities arising from non-generic initial data which are
qualitatively different from those discussed above. They represent an
additional technical hurdle in any attempt to prove cosmic censorship
in general. 

\section{Conclusion}\label{conclusion}

In recent years it has been possible to go beyond the classical 
results on spacetime singularities contained in the singularity
theorems of Penrose and Hawking and close in on the question of
the nature of these singularities in various ways. In the
case of cosmological singularities a key influence has been
exerted by the picture of Belinskii, Khalatnikov and Lifshitz (BKL).  
In the case of black hole singularities the old idea that they 
should be similar to cosmological singularities has been replaced
by the new paradigm of weak null singularities due to Poisson
and Israel. A new kind of singularity has emerged in the work
of Choptuik on critical collapse. It remains to be seen whether
the Einstein equations have further types of singularities in 
store for us.

New things can happen if we go beyond the usual framework of
the singularity theorems. The cosmological acceleration 
which is now well-established by astronomical observations
corresponds on the theoretical level to a violation of the
strong energy condition and suggests that a reworking of
the singularity theorems in a more general context is necessary.
Exotic types of matter which have been introduced to model
accelerated cosmological expansion go even further and violate
the dominant energy condition. This can lead to a cosmological
model running into a singularity when still expanding \cite{starobinsky}. 
This is known as a 'big rip' singularity \cite{caldwell} since physical
systems are ripped apart in finite time as the singularity is approached.
The study of these matters is still in a state of flux. 

Returning to the more conventional setting where the dominant energy
condition is satisfied, we can ask what the future holds for the study of
spacetime singularities in classical general relativity. A
fundamental fact is that our understanding is still very incomplete.
Two developments promise improvements. The first is that the 
steady increase in computing power and improvement of numerical
techniques means that numerical relativity should have big
contributions to make. The second is that advances in the theory
of hyperbolic partial differential equations are providing the tools
needed to make further progress with the mathematical theory of
solutions of the Einstein equations. As illustrated by the examples
of past successes surveyed in this paper the numerical and mathematical
approaches can complement each other very effectively.

\end{document}